\documentclass[twocolumn]{revtex4}
\usepackage{graphicx}
\usepackage{amssymb,amsfonts,amsmath}
\usepackage[usenames]{color}

\usepackage[normalem]{ulem}

\newcommand{\bi}{\begin{itemize}}
\newcommand{\ei}{\end{itemize}}

\newcommand{\kT}{k_{\mathrm{B}}T}
\begin{document}
\title{Irregular transcription dynamics for rapid production of high-fidelity transcripts}
\author{Martin Depken}
\affiliation{Department of Bionanoscience, Kavli Institute of Nanoscience, Delft University of Technology, Delft, The Netherlands}
\affiliation{Department of Physics and Astronomy, Vrije Universiteit, De Boelelaan 1081
1081 HV, Amsterdam, The Netherlands}
\affiliation{Max Planck Institute for the Physics of Complex Systems, N\"othnitzerstrasse
38, 01187 Dresden, Germany}

\author{Juan M. R. Parrondo}
\affiliation{Departamento de F\'{\i}sica At\'omica, Molecular y Nuclear and GISC, Universidad
Complutense de Madrid, 28040-Madrid, Spain}
\affiliation{Max Planck Institute for the Physics of Complex Systems, N\"othnitzerstrasse
38, 01187 Dresden, Germany}

\author{Stephan W. Grill}
\affiliation{Max Planck Institute for the Physics of Complex Systems, N\"othnitzerstrasse
38, 01187 Dresden, Germany}
\affiliation{Max Planck Institute of Molecular Cell Biology and Genetics, Pfotenhauerstr. 108, 01307 Dresden, Germany}

\begin{abstract}
Both genomic stability and sustenance of day-to-day life rely on efficient and accurate readout of the genetic code. Single-molecule experiments show that transcription and replication are highly stochastic and irregular processes, with the polymerases frequently pausing and even reversing direction. While such behavior is recognized as stemming  from a sophisticated proofreading mechanism during replication, the origin and functional significance of irregular transcription dynamics remain controversial. Here, we theoretically examine the implications of RNA polymerase backtracking and transcript cleavage on transcription rates and fidelity. We illustrate how an extended state space for backtracking provides entropic fidelity enhancements that, together with additional fidelity checkpoints, can account for physiological error rates. To explore the competing demands of transcription fidelity, nucleotide triphosphate (NTP) consumption and transcription speed in a physiologically relevant setting, we establish an analytically framework for evaluating transcriptional performance at the level of extended sequences. Using this framework, we reveal a mechanism by which moderately irregular transcription results in astronomical gains in the rate at which extended high-fidelity transcripts can be produced under physiological conditions. 
\end{abstract}


\maketitle

 As organisms evolved and diversified,  more genes, longer genes and bigger genomes needed to be processed~\cite{xu_average_2006}, with increased demands on fidelity. 
Central to fidelity in replication and transcription is that the four different NTPs posses different affinities for pairing with template nucleotides. This results in a preference for forming proper Watson-Crick pairs~\cite{sugimoto_thermodynamic_1995}. Although substantial~\cite{Svetlov:2004dc}, this selectivity is ultimately limited by early and immutable evolutionary choices pertaining to the chemistry of nucleotides. To meet further demands for fidelity, both DNA and RNA polymerases have evolved proofreading mechanisms capable of removing errors which have already been incorporated into their growing polymer product. Through such mechanisms, replication reaches an error ratio (number of incorrect bases divided by the number of correct bases in the final transcript) of the order of $1/10^8$~\cite{kunkel_dna_2004}, while transcription achieves error ratios of the order of $1/10^5$~\cite{sydow_rna_2009}. In this paper we seek to provide a quantitative 
understanding of transcriptional proofreading and its consequences for nucleotide consumption and transcription speed. Due to the incomplete data concerning the microscopic rates for any individual type of polymerase, we here rely on the great structural homology among bacterial, eukaryotic, and archaeal polymerases to~\cite{Ebright:2000,Hirata:2008} to infer {\em order-of-magnitude} estimates of transition rates between microscopic states for a generic polymerase. 

The theoretical underpinning of kinetic proofreading was established by Hopfield over 30 yeas ago~\cite{hopfield_kinetic_1974}. However, the standard treatment assumes the bases to be repeatedly checked {\em before} being {\em permanently} incorporated into the growing transcript. This pre-incorporation selection (PIS) results in an ever growing transcript.  With the event of single-molecule techniques, it is now well established that both RNA and DNA polymerases elongate  their produce in a highly irregular manner: repeatedly pausing, moving backwards, and cleaving bases from the growing molecule~\cite{erie_multiple_1993, donlin_kinetic_1991,thomas_transcriptional_1998,orlova_intrinsic_1995, zenkin_transcript-assisted_2006, wang_structural_2009, galburt_backtracking_2007, wuite_single-molecule_2000, ibarra_proofreading_2009,kireeva_mechanism_2009}. In fact, post-incorporation proofreading (PIP) has long been recognized to play a vital role in error suppression~\cite{erie_multiple_1993, thomas_transcriptional_1998, sydow_rna_2009, ibarra_proofreading_2009, kunkel_dna_2000,jeon_fidelity_1996}, but has received little attention at a quantitative theoretical level~\cite{voliotis_backtracking_2009}. 

We here use stochastic modeling to explore the downstream effects of PIP in transcription, and the connection between proofreading, irregular transcription dynamics, and overall elongation performance. Our stochastic hopping model~\cite{voliotis_backtracking_2009,Greive_Thinking_2005} is built using structurally well characterized states, with transition rates measured in physiologically relevant settings. The model quantitatively couples chain elongation to the observed depolymerizing action of proofreading~\cite{erie_multiple_1993,thomas_transcriptional_1998}.  Through this we show that the highest error-suppression calculated within a standard Hopfield scheme corresponds to a pathological situation with a net shortening of the transcript over time---a fact previously overlooked. This highlights the importance of moving beyond considerations of fidelity alone if we are to gain even a qualitative understanding of this fundamental process.

Proofreading must be efficient on a wide variety of genes, and we adopt a sequence-averaged view to identify a mechanism that works on generic sequences. Through this we are  able to separate the dynamically generated heterogeneity from that of a static, sequence based, origin. We show that the dynamics of an \emph{efficiently} transcribing polymerase should be expected to be irregular---even before taking sequence effects into account. This suggests that a substantial part of the heterogeneous dynamics seen in single-molecule experiments is functionally advantageous and important for ensuring fidelity ~\cite{erie_multiple_1993,thomas_transcriptional_1998,orlova_intrinsic_1995,zenkin_transcript-assisted_2006,wang_structural_2009,shaevitz_backtracking_2003,sydow_rna_2009}. 

\section{Modeling error suppression through PIS and PIP}
Thermal fluctuations are significant on the molecular scale, and we describe transcription as a stochastic hopping process between well defined states, with transition rates set by the intervening free-energy barriers~\cite{risken_fokker-planck_1996}. Following Hopfield~\cite{hopfield_kinetic_1974}, we take the error suppression to be achieved through a sequence of serially connected energy-consuming, molecular-scale, and error-correcting checkpoints. The quality of a checkpoint is judged by its error fraction $r$,  and the quality of several sequential checkpoints is given by the product of individual error fractions $r_1\cdot r_2\cdot r_3\cdot \ldots$ (see supplemental information). 

Error suppression in transcription involves several checkpoints, divided into two classes: PIS and PIP~\cite{sydow_rna_2009}. Contrary to the situation for the DNA polymerase, both types of checkpoints  are controlled by the same multifunctional active region inside the RNA polymerase (RNAP)~\cite{kettenberger_architecture_2003, opalka_structure_2003}. The PIS process likely involves several steps~\cite{sydow_rna_2009} before the incoming NTP establishes the correct Watson-Crick base pairing with the DNA template, and catalyzes onto the growing RNA molecule~\cite{sydow_rna_2009, cramer_gene_2007}.  As the states  prior to catalysis are limited by  the free-energy cost $\Delta G_{\rm act}$ of binding the wrong base to the template DNA strand within the polymerase, $r_{\rm PIS}\ge \exp(-\Delta G_{\rm act}/\kT)$. From direct nucleotide discrimination studies $r_{\rm PIS}$ has been shown to be  $1/10^3-1/10^2$~\cite{Svetlov:2004dc}, corresponding to an average $\Delta G_{\rm act}\approx 6\kT$. Utilizing PIS alone, sequences of no more then a few hundred base pairs (bp) can be reliably transcribed without errors.

To increase fidelity past $r_{\rm PIS}$, and be able to faithfully transcribe longer genes, RNAP has evolved the ability to proofread the transcript by selectively removing already incorporated bases~\cite{sydow_rna_2009,thomas_transcriptional_1998, erie_multiple_1993}. The succesive action of both PIS and PIP is known to bring the combined error fraction $r_{\rm PIS}r_{\rm PIP}$ down to around $1/10^{5}$~\cite{rosenberger_frequency_1983, blank_rna_1986, mercoyrol_accuracy_1992}.  From the estimates of the PIS efficiency mentioned above, we expect half of the error suppression to reside in PIP: $r_{\rm PIP}=1/10^3-1/10^2$. Lead by experimental results we now set out to quantitatively explain how this is achieved in a physiologically relevant setting through the use of extended, backtracked pauses. To highlight the benefits and implications of an extended backtracked state space, we first consider the case of only one backtracked state, and later contrast it to the case with the physiologically more relevant case of multiple states.

\subsection{Proofreading through backtracking}
\begin{figure}[htb!]
\begin{center}
\includegraphics[width=\columnwidth]{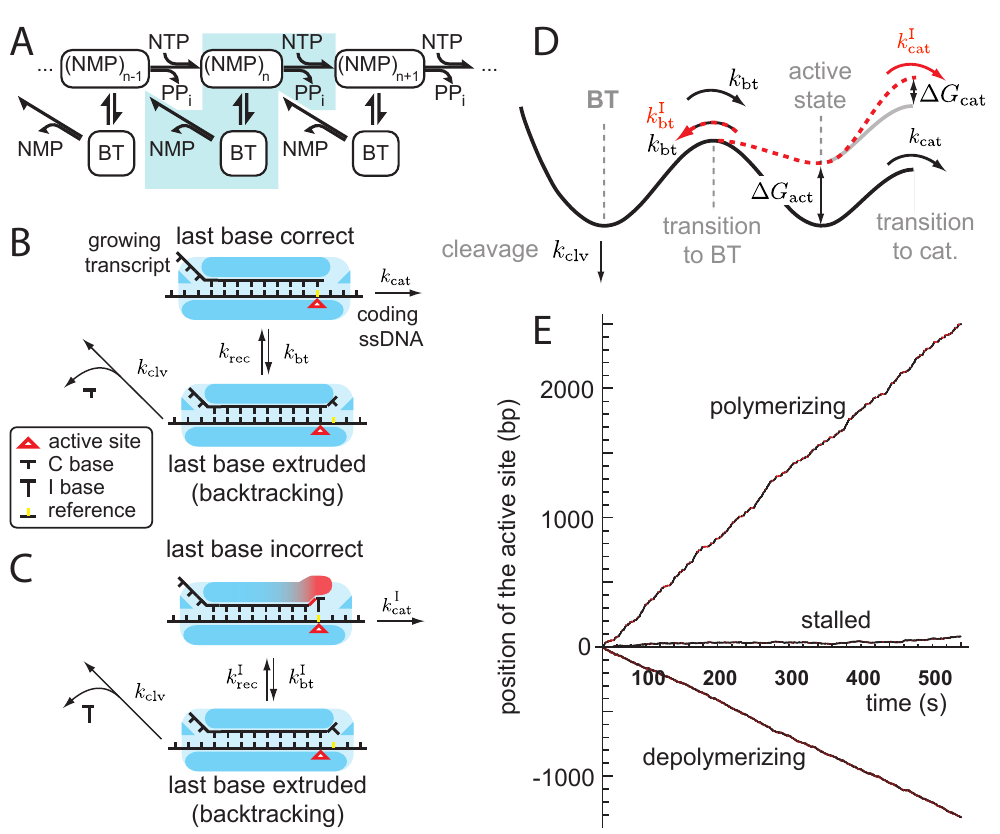}
\end{center}
\caption{\label{fig:back}  {\bf Single-state backtracking}. A) The basic hopping model coupling one-step backtracking to elongation.  The repetitive unit is highlighted, with the off-pathway backtracked state indicated as BT. After entering a backtrack, elongation can resume either through cleavage out to a previous state of the chain (NMP)$_{n-1}$ or by recovery without cleavage to the entrance state (NMP)$_{n}$.  B) Schematic illustration of the repeat unit with a correct base incorporated last. The template strand, the nascent transcript, and the hybrid region of the polymerase are shown.  The polymerase can enter a backtrack with rate $k_{\rm bt}$ or add a base to the transcript with rate $k_{\rm cat}$. From the backtracked state, recovery by cleavage occurs with rate $k_{\rm clv}$, while realigning without cleavage occurs at a rate $k_{\rm rec}$. C) Same as B, but with an incorrect base at the growing 3'-end of the transcript. The  corresponding rates are indicated with the superscript ${\rm I}$. D) Sketch of the free-energy landscape corresponding to B and C. Solid black line corresponds to the last base correct; dashed red line corresponds to the last base incorrect. $\Delta G_{\rm act}$ refers to the free-energy increase at the active site when the last incorporated base is wrong, while $\Delta G_{\rm cat}$ denotes the corresponding increase in the barrier to catalysis (cat). Recovery without cleavage occurs at a rate $k_{\rm bt}$, which places all selectivity in the entrance step to the backtrack (see text). E) Three traces simulated with a Gillespie algorithm: a typical polymerizing RNAP with ($k_{\rm cat}=10/$s, $k_{\rm bt}=1/$s, $k_{\rm clv}=0.1/$s, see main  text), a stalled polymerase ($k_{\rm cat}=1/$s, $k_{\rm bt}=10/9$s, $k_{\rm clv}=10/$s), and a depolymerase ($k_{\rm cat}=1/$s, $k_{\rm bt}=10/$s, $k_{\rm clv}=10/$s). Traces are black when the polymerase is elongating, and red when backtracked.}
\end{figure}

It is well established that an erroneous base can be cleaved from the growing transcript once the polymerase has entered what is known as a backtracked state\footnote{There is some evidence in the literature for an intermediate state between elongation and backtracking~\cite{herbert_E._2010}. However, the rates for transversing this state are similar to those for entering the backtrack, and adding such a state does not change the general dynamics of the model.} (see Figure~\ref{fig:back}A): an off pathway state where the whole polymerase is displaced backward along the transcript~\cite{shaevitz_backtracking_2003, galburt_backtracking_2007}. Within the polymerase, the template DNA and nascent RNA strands form a 8-9 bp hybrid. As the polymerase shifts backward, this hybrid remains in register by breaking the last formed bond and reforming an old bond at the opposite end of the hybrid~\cite{shaevitz_backtracking_2003, galburt_backtracking_2007,wang_structural_2009} (see Figure~\ref{fig:back}B and C). This exposes already incorporated bases to the active site, blocking further elongation but enabling cleavage of the most recently added base (catalyzed by the transcription factor IIS in eukaryotes and GreA and GreB in prokaryotes)~\cite{fish_promoting_2002,borukhov_bacterial_2005,jeon_fidelity_1996,awrey_yeast_1998,opalka_structure_2003, kettenberger_architecture_2003, sosunov_unified_2003, thomas_transcriptional_1998}. If cleaved, a potential error is removed, the active site is cleared, and elongation can resume. The cleavage process competes with the spontaneous recovery from the backtrack~\cite{galburt_backtracking_2007}, by which the polymerase returns to the elongation competent state without removing the potential error (see Figure~\ref{fig:back}C).  In order for cleavage from the backtracked state to lower the error content, the cleavage reaction must select for erroneous bases. The inability of incorrectly matched bases to form proper Watson-Crick base paring within the RNA-DNA hybrid induces this selectivity. If an error has been catalyzed onto the 3'-end of the nascent RNA molecule, the total energy of the transcription complex is lowered if the RNAP moves into a backtrack (see Figure ~\ref{fig:back}D). Doing this, the RNAP extrudes the unmatched
base pair from the hybrid and so returns to the low energy state of a perfect Watson-Crick  base-pairing within the entire hybrid (see Figure ~\ref{fig:back}C). When the polymerase is in a backtracked state, the last added base is exposed to the active site and can be cleaved off.

How much cleavage from backtracked states contributes to error suppression depends on the effect of misincorporations on the transition rates in and out of backtracks. Specifically, the manner in which a missincorporation effects the transition state to backtracking determines if fidelity increases will be affected through an increased entrance rate into the backtrack (no shift of transition state) or a lowered exit rate out of the backtrack (transition state shifts with the hybrid energy). For the latter case to have an appreciable proofreading capability, every single base must at some point be extruded out of the polymerase through backtracking, such that the base can be proofread and removed if it happen to be incorrectly matched to the template strand. The required high backtracking frequency would render the polymerization process inefficient---even reverse it (see below)---which is clearly not what is observed in experiments~\cite{galburt_backtracking_2007,abbondanzieri_direct_2005}. We thus take the selectivity to reside in the entrance step of the backtrack (see Figure~\ref{fig:back}D). For rates as illustrated in Figure~\ref{fig:back}B and C, this corresponds to $k_{\rm rec}=k_{\rm rec}^{\rm I}=k_{\rm bt}$ and $k_{\rm bt}^{\rm I}=k_{\rm bt}\exp(\Delta G_{\rm act}/\kT)$ (rates corresponding to incorrect bases are denoted with the superscript I). We will simply refer to $k_{\rm bt}$ as the backtracking rate, and the resulting form of the free-energy landscape is illustrated in Figure~\ref{fig:back}D.

\subsection{Physiological rate estimates}
Although single-molecule traces give us direct access to many of the individual rates introduced in Figure~\ref{fig:back}B and C,  the spread even between individual enzymes of any specific type of polymerase is substantial~\cite{neuman_ubiquitous_2003, toli-nrrelykke_diversity_2004}. On top of this, not all rates are known for any one type of polymerase, so we are here content with relying on the structural homology between polymerases ~\cite{Ebright:2000,Hirata:2008} and take  {\it in vitro} rates from the different domains as representing order of magnitude estimates of a generic enzyme. We use $k_{{\rm cat}}=10$/s~\cite{neuman_ubiquitous_2003,toli-nrrelykke_diversity_2004} (prokaryotic) \cite{galburt_backtracking_2007} (eukaryotic), backtracking rate $k_{{\rm bt}}=1$/s~\cite{depken_origin_2009} (prokaryotic), and cleavage rate $k_{{\rm clv}}=0.1$/s \cite{galburt_backtracking_2007} (eukaryotic). Though this will not cover every scenario, the analytical nature of our work enables direct application of our results to other relevant situations. 

In a development largely parallel to the theory of kinetic proofreading through PIS~\cite{hopfield_kinetic_1974}, the error suppression of PIP can be calculated as (see supplemental information)
\begin{equation}
\label{eq:pip0}
r\simeq \frac{k_{\rm cat}}{k_{\rm cat}+k_{\rm clv} e^{(\Delta G_{\rm act}+\Delta G_{\rm cat})/\kT}}.
\end{equation}
Here $\Delta G_{\rm cat}$ denotes the change in barrier height for the transition to catalysis when trying to incorporate a base directly after an error (see Figure~\ref{fig:back}D). We can get an estimate of $\Delta G_{\rm cat}$ from published experiments that use "non-hydrolizable" nucleotide substitutes. These substitutes are thought not to influence binding affinities, but to change the catalysis rate to an extent comparable to that of an erroneous base~ \cite{thomas_transcriptional_1998}. From this we estimate $\Delta G_{\rm cat}\approx 2\kT$. For our typical polymerase this implies proofreading capabilities amounting to a modest $r_{\rm PIP}\approx 1/30$: off by an order of magnitude from the experimentally determined fidelity ($1/10^3-1/10^2$).  Note that the error ratio is insensitive to $k_{\rm bt}$ for our typical polymerase ($k_{\rm cat} \gg k_{\rm bt} \gg k_{\rm clv}$). Further, a comparison of the regular traces (see Figure~\ref{fig:back}E) resulting from this model (see Figure~\ref{fig:back}A) with those from single-molecule experiments~\cite{galburt_backtracking_2007,depken_origin_2009, ibarra_proofreading_2009} demonstrates that the model does not adequately capture the observed irregular transcription dynamics (see also below). Although much of the observed dynamical heterogeneity has been attributed to structural heterogeneity through sequence specific pauses~\cite{herbert_sequence-resolved_2006,neuman_ubiquitous_2003, ibarra_proofreading_2009}, we here show that this is not {\em necessarily} the dominant contribution.

\subsection{Entropic fidelity enhancements}
It is clear from Equation~\ref{eq:pip0} that apart from increasing the energy penalty for a bad basepair, a low error-ratio can be achieved through a relative increase of the transcript cleavage rate compared to the elongation rate. Given their reverse arrangement ($k_{\rm cat}\simeq 10/{\rm s} \gg k_{\rm clv}\simeq 0.1/{\rm s}$), we speculate that the evolution of these rates has been strongly limited by external constraints pertaining to nucleotide chemistry and the intercellular environment. To mediate these external constraints, the polymerase has had to find alternative {\em internal} paths to increase error suppression. 

\begin{figure}[htb!]
\begin{center}\includegraphics[width=\columnwidth]{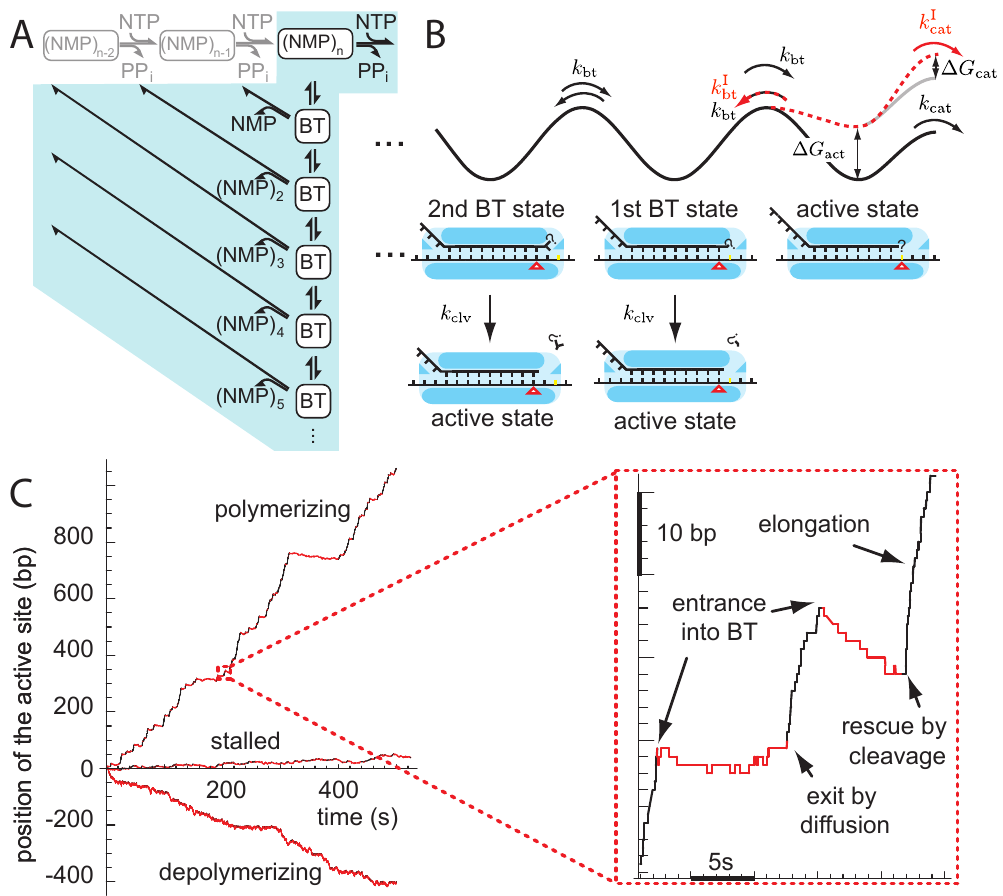}\end{center}
\caption{\label{fig:entropy}  {\bf Multi-state backtracking.} 
A) The basic repeat unit of multi-state backtracking in a nested scheme. For visual clarity, only the backtracked states in the highlighted repeat unit are drawn. B) Sketch of the free-energy landscape of a multi-state backtrack. Solid black line corresponds to the last base correct; dashed red line corresponds to the last base incorrect. Also illustrated are the multiple backtracked states and the effect of cleavage. See caption to Figure~\ref{fig:back}B for a description of the rates. C) Three traces simulated with a Gillespie algorithm: a typical polymerizing RNAP with ($k_{\rm cat}=10$, $k_{\rm bt}=1$, $k_{\rm clv}=0.1$), a stalled complex ($k_{\rm cat}=10$, $k_{\rm bt}=10$, $k_{\rm clv}=0.1$), and a depolymerizing one ($k_{\rm cat}=1$, $k_{\rm bt}=10$, $k_{\rm clv}=0.1$)---all in accordance to the theoretical predictions derived in the supplemental information. A section of the trace for our typical polymerase has been magnified, showing two backtracks, one rescued to elongation by cleavage and one by diffusion. Only the backtrack reentering elongation through cleavage would have corrected an error at the end of the transcript.  Traces are black when the polymerase is elongating, and red when backtracked.}
\end{figure}

One such internal path could be to reduce the free energy of the backtracked state. This would suppress spontaneous reversal of the backtrack and therethrough increase the probability of cleavage and error removal. Since a substantial part of the free energy relates to the energetics of base matching within the hybrid,  the energy level of the backtracked state is likely constrained by the structure of the hybrid---again presumably fixed by early evolutionary choices. 
However, nature appears to have come up with a different solution: an effective entropic reduction in the free-energy level of the backtracked state is achieved by extending the number of accessible states. RNAP is able to backtrack by more than just one base, and thermally move between the different backtracking states that are available~\cite{nudler_rna_1997, komissarova_rna_1997, komissarova_transcriptional_1997,shaevitz_backtracking_2003,galburt_backtracking_2007} (see Figure~\ref{fig:entropy}A).  With $N$ off-pathway and backtracked proofreading states, the free energy associated with the backtracked state would, in an equilibrium setting, be reduced by the entropic term $\kT \ln(N)$. Even in our out of equilibrium setting this mechanism delays spontaneous recovery and raises the chance of cleavage and error removal (see supplemental information). 
With an extended backtracking space\footnote{Even with infinite room for backtracking, our typical polymerase would only take around $k_{\rm bt}/k_{\rm clv}= 10$ diffusive backtracking steps before being cleaved off, and would reach a typical backtracking depth of around $N\approx \sqrt{k_{\rm bt}/k_{\rm clv}}\approx 3$. This is below the lower estimates for the distance to RNA hairpin barriers in the trailing RNA strand~\cite{klopper_influence_2010}. We are thus justified in assuming the available backtracking distance to be effectively infinite (see Figure~\ref{fig:entropy}A). }, it is now clear from simulated traces (Figure~\ref{fig:entropy}C)  that the irregular dynamics of our typical polymerase qualitatively matches the irregular dynamics observed in single-molecule experiments~\cite{herbert_sequence-resolved_2006,neuman_ubiquitous_2003, ibarra_proofreading_2009} (see below for a quantitative assessment).  By comparing the experimental effects of cleavage stimulating factors and simulated traces for increased cleavage rates, we provide further support of our kinetic scheme in the supplemental information. We also show that our model can capture the stalling dynamics of a polymerase as it transcribes against an increasing force~\cite{galburt_backtracking_2007}.

When acting through extended backtracked states, the error suppression of PIP can be calculated as (see supplemental information)
\begin{eqnarray}
\label{eq:r}
 r_{\rm 1:PIP}&\simeq&\frac{k_{\rm cat}}{k_{\rm cat}+\sqrt{k_{\rm clv}k_{\rm bt}} e^{(\Delta G_{\rm act}+\Delta G_{\rm cat})/\kT}}\\
&=& \frac{k_{\rm cat}}{k_{\rm cat}+k_{\rm bt} e^{\Delta G_{\rm 1:PIP}/\kT}},\nonumber\\
\Delta G_{\rm 1:PIP}&=&\Delta G_{\rm act}+\Delta G_{\rm cat}-\frac{1}{2}\kT\ln(k_{\rm bt}/k_{\rm clv}).\label{eq:r2}
\end{eqnarray}

Comparing Equation~\ref{eq:r} to Equation~\ref{eq:pip0}, we see that 
 fidelity is increased by extending the space available for backtracking: the low cleavage rate $k_{\rm clv}$ is replaced by the geometric mean  $\sqrt{k_{\rm clv}k_{\rm bt}}$. This increases the fidelity by about a factor of three for our typical polymerase, and provides an error reduction of $r_{1:\rm PIP}\simeq 1/100$. The notation in Equation~\ref{eq:r2} is introduced to facilitate the extension to several PIP checkpoints presented in the next section. The error suppression now depends on the additional parameter $k_{\rm bt}$ (c.f. Equation~\ref{eq:pip0})---a parameter independent of nucleotide chemistry and susceptible to change through evolutionary pressures.
Although the extension of the backtracking space does provide for fidelity enhancements, the total fidelity is still at the lower end of what is experimentally observed. However, our extended backtracking space gives further proofreading benefits by supplying the polymerase with additional inherent PIP checkpoints, as we now discuss.
\begin{figure}[htb!]
\begin{center}
\includegraphics[width=\columnwidth]{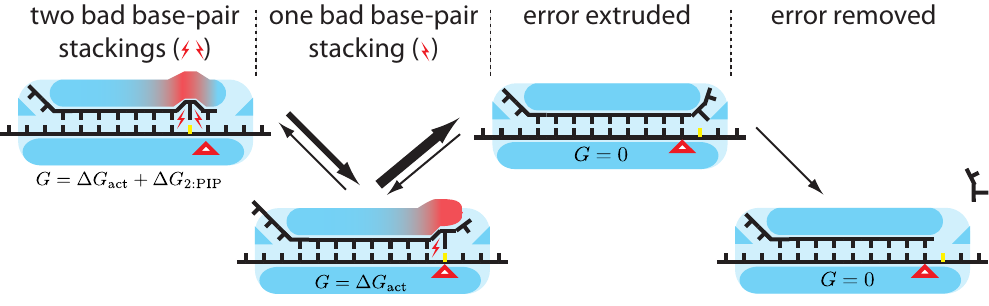}
\end{center}
\caption{\label{fig:CPIII}  {\bf A second PIP checkpoint.} The polymerase is expected to be sensitive to errors incorporated also next to last. The magnitude of the rates are illustrated by relative thickness of the transition arrows, bad base stackings are indicated in red.  $G$ indicates the free energy of the complex with respect to the elongation competent state.}
\end{figure}
\subsection{Second PIP checkpoint and beyond}
Even when additional bases have been added to the transcript after an erroneous incorporation, the error can in principle still be corrected through an extensive backtrack and cleavage~\cite{voliotis_backtracking_2009}. For this to lead to an appreciably increased likelihood of error removal, the random walk  must be biased towards entering further into the backtrack. With an error at the {\em penultimate} 3'-position of the transcript, the polymerase experiences such bias, since moving into a backtrack will eliminate a bad base-pair stacking within the hybrid (see Figure~\ref{fig:CPIII}). This is followed by another heavily biased step to completely extrude the error from the hybrid, making it amenable to cleavage.  We know of no direct measurement of the penultimate bias $\Delta G_{\rm 2:PIP}$, but as the typical stacking energy in a RNA-DNA hybrid is $1.5-4.5\,\kT$~\cite{sugimoto_thermodynamic_1995} we assume $\Delta G_{\rm 2:PIP}\approx 3\kT$.  This second PIP checkpoint provides an error ratio (see supplemental information) of
\begin{equation}\label{eq:rp}
r_{2:\rm PIP}\simeq\frac{k_{\rm cat}}{k_{\rm cat}+k_{\rm bt} e^{\Delta G_{\rm 2:PIP}/\kT}}.
\end{equation}
For our typical polymerase $r_{2:\rm PIP}\simeq 1/3$, and the total PIP induced error reduction $r_{\rm PIP}=r_{1:\rm PIP} r_{2:\rm PIP}\simeq 1/300$ falls well within the experimentally observed range. 

The suggested scheme thus quantitatively accounts for the typically observed error-suppression, but there could in principle be 
additional inherent PIP checkpoints that would enable the polymerase to reach even higher fidelities. An increasing free-energy penalty for moving the error further into the hybrid would incur a longer range bias for backtracking, and additional fidelity gains according to (see supplemental information)
\begin{eqnarray}\label{eq:mp}
r_{\rm PIP} &=& r_{1:\rm PIP}r_{2:\rm PIP}\cdots r_{n\rm :PIP}\cdots\quad\nonumber \\
r_{n:\rm PIP}&\simeq&\frac{k_{\rm cat}}{k_{\rm cat}+k_{\rm bt} e^{\Delta G_{n\rm :PIP}/\kT}}.
\end{eqnarray}
Based on structural considerations of base pairing within the RNA-DNA hybrid, we conclude that PIP-proofreading of RNAP includes at least two serial checkpoints that account for the typical fidelities observed in transcription. The polymerase could in principle select and remove an error as long as it remains within the hybrid. Intriguingly,  the 8-9-bp hybrid might thus not only serve the purpose of stabilizing the ternary complex~\cite{nudler_rna-dna_1997}, but also provide enhanced fidelity.

\subsection{Power-law pause distributions and spatial heterogeneity}
\begin{figure}[htb!]
\begin{center}
\includegraphics[width=\columnwidth]{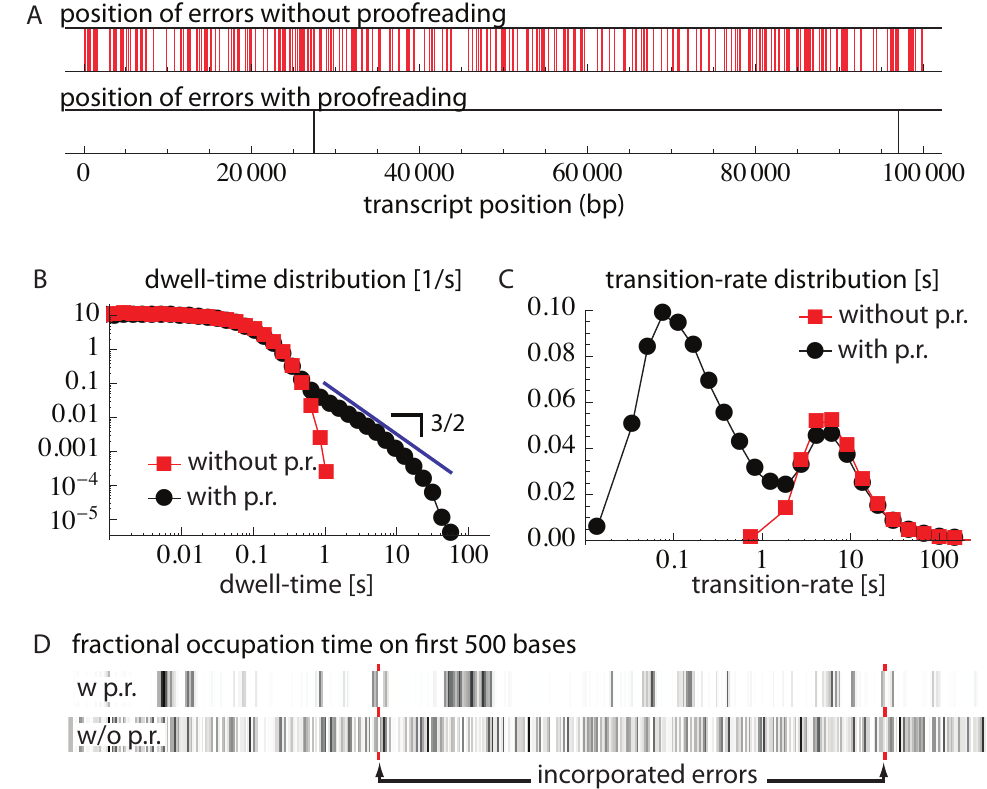}
\end{center}
\caption{\label{fig:sim}  {\bf The effects of proofreading.} A) On top we show a realization of incorporation errors according to our free-energy estimates (only PIS in red), and below we show the errors that survive (or, possibly, are inserted by) the proofreading mechanisms (PIS and PIP in black). B) The dwell-time distribution from a process without proofreading and one with proofreading. Proofreading gives rise to a power-law regime significantly increasing the fraction of long-pauses. C) The transition-rate (inverse dwell-time) distribution for the same processes as in B, where the effects of proofreading can be seen through a shift from a unimodal to a bimodal distribution as many excessively slow transitions involving backtracks start influencing the kinetics.  D) The pause density, or the total occupation time plotted for a 500 bp sequence transcribed by the same two polymerases as used in B and C. The darker the bands,  the longer the total occupation time at that position. The scales are individually normalized to cover the range of occupation times for each polymerase. Two incorporation errors are indicated with red markers.}
\end{figure}

We next illustrate the consequences of the proofreading states on pause duration and frequency. 
 To this end we simulate our typical polymerase transcribing a  long sequence and compare it to a simulation of an otherwise identical polymerase, but which has PIP turned off ($k_{\rm bt}=0$/s). In Figure~\ref{fig:sim}A we show a particular realization (of our generic polymerase) of incorporation errors (only PIS in red) together with the errors left after the section has been proofread (PIS and PIP in black). The fidelity enhancements are clearly visible, but they come at the cost of both an decreased velocity, as well as an increased spatial heterogeneity.  These effects are qualitatively visible already at the level of individual traces, but are quantitatively best seen in the changes of the dwell-time distribution (see Figure~\ref{fig:sim}B) or in the transition-rate (inverse dwell-time) distribution (see Figure~\ref{fig:sim}C). In the dwell-time distribution, proofreading introduces a  power-law regime, throughout  which the probability of a long pause falls off with duration $t$ as $t^{-3/2}$~\cite{depken_origin_2009}, until it drops off exponentially beyond $t\sim 1/k_{\rm clv}$. In Figure~\ref{fig:sim}B we see a clear exponential behavior of the dwell-time distribution for both processes at around $t\sim 1/k_{\rm el}=0.1$s, while the proofreading polymerase also has the above mentioned  power-law decay extending out to $t\sim 1/k_{\rm clv}=10$s.  Similarly, considering the transition-rate distributions we see a narrow but significant low velocity peak develop around the transition rate$\sim k_{\rm clv}=0.1/$s, diminishing the bare elongation peak situated around the rate$\sim k_{\rm cat}=10/$s (see Figure~\ref{fig:sim}C). To further elucidate the effects of the power-law regime, we consider another important observable: the pause-time distribution, or the total time a polymerase spends at each position along the DNA molecule. In Figure~\ref{fig:sim}D we show pause density plots along a sequence of 500 bp, with darker bands indicating longer total time spent at that position during the transcription process. Comparing transcription with and without PIP it is clear that PIP leads to greater spatial heterogeneity, exhibiting distinct regions of markedly increased occupation density even where there are no incorporation errors. Thus, our model accounts for both the observed spatial heterogeneity as well as the broad pause-time distributions~\cite{neuman_ubiquitous_2003,galburt_backtracking_2007} without the need to introduce additional assumptions about the effects of sequence heterogeneity~\cite{depken_origin_2009, galburt_backtracking_2007}. 

Having shown that external constraints can be mediated through accessing an internal extended backtracked space---resulting in irregular transcription dynamics---we now turn our attention to the specific level of irregularity observed in experiments. Irregularity is tuned by the backtracking rate, and considering that increasing $k_{\rm bt}$ would render all proofreading checkpoints more effective (see Equation~\ref{eq:mp}) one might wonder why the backtracking rate is kept moderate (1/s) and not made much larger~\cite{depken_origin_2009,abbondanzieri_direct_2005}.

\section{Transcription performance}
We have here suggested that by utilizing {\em extended} backtracked states, the polymerase has overcome external constraints to suppress errors. This introduces the backtracking rate $k_{\rm bt}$ as a variable susceptible to evolutionary pressures. In order to understand the underlying reasons for why the backtracking rate is kept moderate, we now consider the phenotypic space made available  through the extended backtracking space. The quantities needed to access polymerase performance---as it varies with the level of PIP---are calculated in the supplemental information by using continuous time random walk theory~\cite{montroll_fluctuation_1987}. Starting with instantaneous transcriptional efficiency measures on the level of the individual base pairs, we then consider the efficiency on extended sequences or genes. Importantly, we investigate how much faster the polymerase can produce perfect transcripts of extended sequences with PIP as compared to without PIP.

\subsection{Performance on the level of a base pair}
We are interested in the effective elongation rate, and thus calculate the average elongation rate $1/\tau_{\rm el}$ (see supplemental information). Since there is only about one error passing through the PIS checkpoint every 500 bases, we can ignore the effect of errors on the overall elongation dynamics. We now construct the efficiency measure $\eta_{\rm el}$,
\begin{equation}
\label{eq:tau}
\eta_{\rm el}=\frac{1/\tau_{\rm el}}{k_{\rm cat}}\simeq \frac{1-k_{\rm bt}/k_{\rm cat}}{1/2+\sqrt{1/4+k_{\rm bt}/k_{\rm clv}}},
\end{equation}
which describes the relative slowdown due to PIP. With no PIP ($k_{\rm bt}=0$) the efficiency is appropriately $\eta_{\rm el}=1$, while it vanishes at the transition between polymerization and depolymerization $ k_{\rm cat}=k_{\rm bt}$. At this point, elongation stops proceeding with a well defined velocity, and behaves diffusively on large lengthscales. For $k_{\rm cat}<k_{\rm bt}$ net depolymerization sets in.  
This situation is pathological, and shows that backtracking cannot dominate the dynamics even though this would be judged optimal in terms of fidelity calculated within the Hopfield kinetic proofreading scheme.  The transition to non-functional polymerases can be seen in the single-molecule transcription traces presented in\footnote{See Figure 3D in~\cite{galburt_backtracking_2007}, where an opposing force was used to increase the entrance rate into the backtrack, bringing the system to stall around 14pN.}~\cite{galburt_backtracking_2007}, and in the simulated traces presented in Figure~\ref{fig:entropy}C (see also supplemental Figure S4C). Also note that the overall elongation rate increases with increasing cleavage rate, as is observed experimentally~\cite{fish_promoting_2002, herbert_E._2010,proshkin_cooperation_2010}. We next introduce an efficiency parameter for PIP, 
\begin{equation*}
\eta_{\rm PIP}=1-r_{\rm PIP},
\end{equation*}
which is $0$ in the absence of PIP and $1$ for perfect PIP. Finally, we parameterize the nucleotide efficiency of the transcription process by the ratio of final transcript length and the average number of nucleotides consumed in its production. This ratio is given by the simple expression (see supplemental information)
\begin{equation*}
\eta_{\rm NTP}=1-k_{\rm bt}/k_{\rm cat}.
\end{equation*}
The measure is unity without PIP, and vanishes at stall ($\eta_{\rm el} = 0$).

\begin{figure}[htb!]
\begin{center}
\includegraphics[width=\columnwidth]{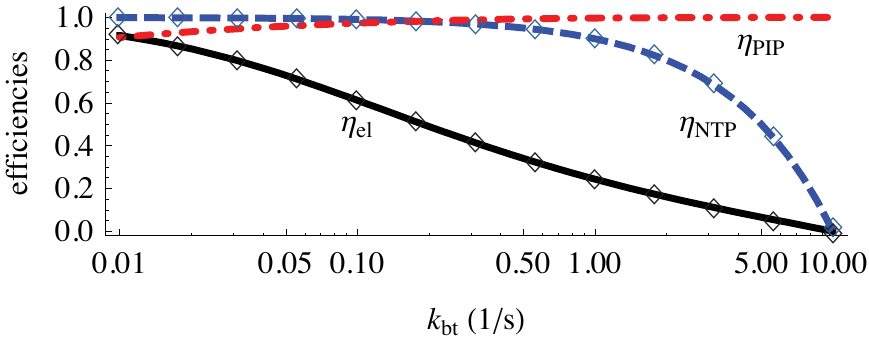}
\end{center}
 \caption{\label{fig:evo} {\bf Polymerase performance.} Proofreading efficiency $\eta_{\rm PIP}$ (red dot-dashed), elongation efficiency $\eta_{\rm el}$ (black solid) and nucleotide efficiency $\eta_{\rm NTP}$ (blue dashed) as a function of the backtracking rate, for an otherwise typical polymerase with $k_{\rm cat}=10$/s and $k_{\rm clv}=0.1$/s. Values indicated by diamonds were obtained numerically, through Gillespie simulations.}
\end{figure}

Figure \ref{fig:evo} shows the three efficiency measures $\eta_{\rm el}$, $\eta_{\rm PIP}$ and $\eta_{\rm NTP}$ as functions of the backtracking rate $k_{\rm bt}$ (within the operational range $0\le k_{\rm bt}\le k_{\rm cat}\approx10$), for an otherwise typical polymerase. We see that while transcription velocity and nucleotide efficiency correlate positively, they both correlate negatively with fidelity, directly illustrating the cost of enhancing fidelity. This hints at an underlying  competition, 
which we now explore by considering transcription of extended sequences.  
\subsection{Performance on the level of the gene}
Here we demonstrate that a moderate rate of backtracking is necessary for rapidly generating transcripts with few mistakes from extended sequences. This becomes apparent when noting that the longer the sequence, the less likely it is for a  polymerase to produce an error-free transcript.  It is instructive to introduce the probability $P_l$ of producing a long error-free sequence\footnote{This sequence length $l$ should not necessarily be interpreted as the complete gene length $l_{\rm gene}$, but instead as the typical  error-free length $l=l_{\rm gene}/n$ that is required, where $n$ is the number of errors acceptable during transcription of the gene.} of length $l$.
For each attempt, the probability of transcribing  a sequence of length $l$ without an error is given by $P_l(r)=(1+r)^{-l}\simeq\exp(-l r)$,
with $r=r_{\rm PIP}r_{\rm PIS}$ representing the total error fraction.
The production-rate gain $\chi_{\rm el}$ on extended sequences is obtained by
comparing the rate at which error-free transcripts are produced with
PIP, to the rate with which they are produced without PIP ($k_{\rm
bt}=0$). Thus, $\chi_{\rm el}=\eta_{\rm el}P_l\left(r_{\rm
PIS}r_{\rm PIP}\right)/P_l\left(r_{\rm PIS}\right)\simeq \eta_{\rm
el} \exp(lr_{\rm PIS}\eta_{\rm PIP})$. Similarly, we introduce the
NTP-efficiency gain on extended genes $\chi_{\rm NTP}$ by
comparing the number of error-free transcripts produced per
nucleotide used with and without PIP, giving $\chi_{\rm
NTP}=\eta_{\rm NTP} P_l\left(r_{\rm PIS}r_{\rm
PIP}\right)/P_l\left(r_{\rm PIS}\right)\simeq \eta_{\rm
NTP}\exp(lr_{\rm PIS}\eta_{\rm PIP})$. From both these quantities it
is clear that even moderate PIP provides enormous gains in the rate of perfectly transcribing long ($l>1/r_{\rm PIS}$) sequences.  
\begin{figure}[htb!]
\begin{center}
\includegraphics[width=\columnwidth]{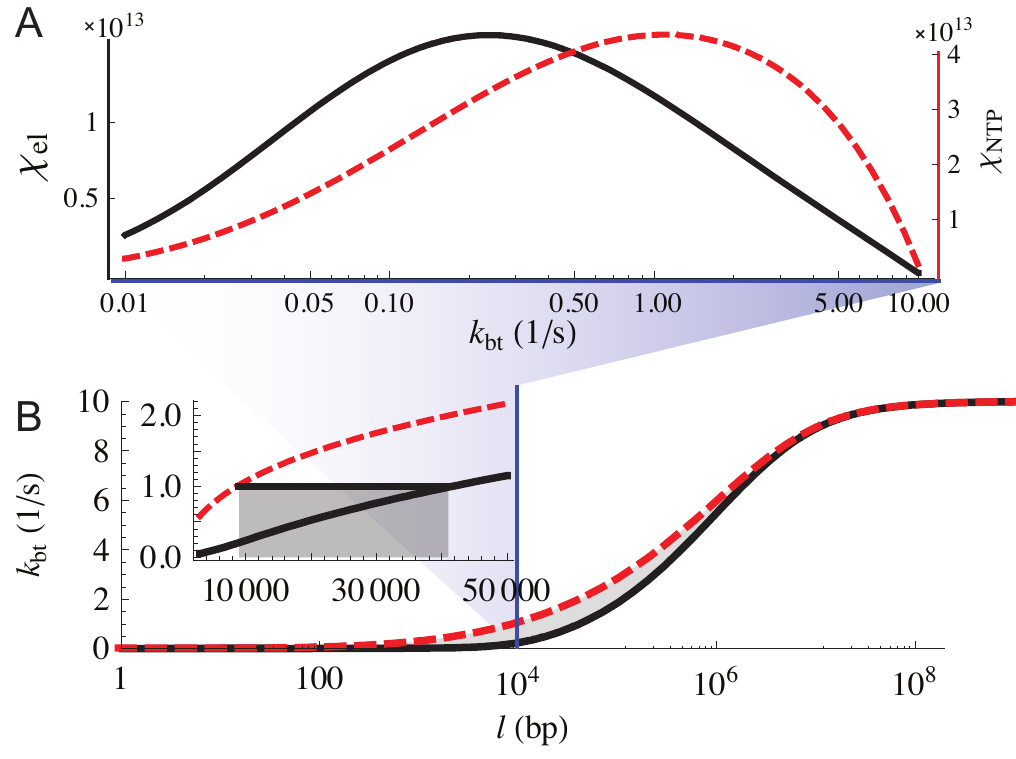}
\end{center}
 \caption{\label{fig:evon} {\bf High fidelity transcript production.} A)  On the left vertical axis we mark the production-rate gain on extended sequences $\chi_{\rm el}$ as a function of the backtracking rate (black solid line). On the right vertical axis we mark the NTP-efficiency gain $\chi_{\rm NTP}$  as a function of the backtracking rate (red dashed line), all for a sequence of length $l=10^4$ bp.  The region between the two peaks is where one might expect the optimal value of $k_{\rm bt}$ to lie. Note there is a gain of 13 orders of magnitude in the rate of producing error-free transcripts when transcribing with PIP as compared to without PIP, with similar gains in nucleotide efficiency. B) The backtracking rate that optimizes the production-rate gain (black solid) or the energy-efficiency gain (red dashed) as a function of  sequence length.  Gray shading indicates a region of compromise between both gains. Inset,  a magnification of the region around $k_{\rm bt}=1/$s indicates that PIP is optimal with $k_{\rm bt}=1/$s for gene lengths of $10^4-4 \cdot10^4$~bp. The vertical blue line indicates the sequence length used in A. }
\end{figure}
With the two sequence-wide measures that we have introduced, it is now possible to address transcriptional efficiencies on the level of transcription of whole genes. As an example we consider a sequence of a length comparable to the typical human gene $l=10^{4}$ bp, and in Figure \ref{fig:evon}A we plot the efficiencies $\chi_{\rm el}$ and $\chi_{\rm NTP}$ as a function of the backtracking rate $k_{\rm bt}$ (within the operational limits $0<k_{\rm bt}<k_{\rm cat}=10$/s). Each measure has a definite optimal value, and we see that the gains in both rate of perfect transcript production and nucleotide efficiency can be enormous, here reaching thirteen orders of magnitude. If RNAP was optimized to transcribe this particular sequence length,  then we would expect the true value of the backtracking rate to lie somewhere in the intermediate region between the peaks: representing a compromise between NTP efficiency and production rate. For the intimidate value of $k_{\rm bt}=1/$s---coinciding with our estimate of the physiologically relevant backtracking rate---it would take a polymerase of the order of one hour to produce an error free transcript, which should be compared to $10^{13}$ hours without PIP. 

Finally,  it is interesting to ask how the region of optimal backtracking rate changes as the transcribed sequence length varies. Figure \ref{fig:evon}B shows the $k_{\rm bt}$ that optimizes $\chi_{\rm el}$ (black solid line) and $\chi_{\rm NTP}$ (red dashed line) as a function of sequence length $l$. The inset in Figure~\ref{fig:evon}B highlights the backtracking rate for our typical polymerase ($k_{\rm bt}=1$/s), and the implied sequence lengths ($\simeq 10^4-4\cdot 10^4$bp) for which this backtracking rate would be optimal. A complete discussion would need to consider relaxed fidelity constraints due to e.g. codon redundancy~\cite{alberts_essential_1998}, but considering that the average gene length in eukaryotes lies in the range $10^4-10^5$~\cite{xu_average_2006}, it is thought-provoking to speculate that the moderate observed backtracking rates of around $1/$s are the result of an evolutionary optimization for rapidly and efficiently producing functional transcripts from genes in the tens-of-kbp range.  
\section{Discussion}
By analytically studying a model of backtracking couple to chain elongation and cleavage, we have shown that irregular transcription dynamics is likely a result of maintaining transcriptional efficiency, not at the level of individual nucleotides, but rather, at the level of extended sequences and genes.   Our work suggests that proofreading relies on an entropic enhancement of fidelity, where an extended state space reduces the chance of spontaneous recovery. This  ensures low error rates even with low rates of transcript cleavage. Through backtracking, an incorporated error can be proofread at least twice through biasing the entry into backtracks, but could in principle be proofread as many times as there are bases in the RNA-DNA hybrid  within the elongation complex.  To what extent there are additional proofreading checkpoints beyond the two discussed here is an interesting line of future research,  providing a  potential  link between the structure of the elongation complex and overall transcriptional efficiency and fidelity. Such work might offer additional clues as to why the RNA-DNA hybrid has a length of about 8-9 bp~\cite{kent_maintenance_2009}.

Considering both the effects of proofreading on NTP consumption and the production rate of extended functional transcripts, our investigation suggests that the internal hopping rate in the backtracked state is not optimized for fidelity alone. Instead, it is kept moderate in order to enable rapid production of extended transcripts that are of high fidelity. That there will be many more backtracks than there are errors to remove is a direct consequence of undetected errors being costly, since they have the potential to render the whole transcript dysfunctional. A certain level of paranoia is thus desirable on part of the polymerase. Even though such paranoia decreases the instantaneous average transcription rate, the observed level of backtracking---perhaps counterintuitively---drastically increases the rate at which high fidelity transcripts are produced. Interestingly, the backtracking rate and the amount of backtracks in cells of a particular organism would be expected to correlate positively with the sequences length that has induced the highest evolutionary pressures on transcription (see Figure~\ref{fig:evon}B, gray region).  In other words,  genomes with genes of increasing length should be transcribed with increasingly irregular dynamics to maintain transcriptional efficiency. It would be interesting to determine if an overall trend in backtracking rate~\cite{galburt_backtracking_2007, depken_origin_2009}, and consequent irregularity of dynamics, could be found for polymerases originating in organisms with varying genetic complexity.

To conclude, our model highlights the enormous gains offered by post-incorporation proofreading when transcribing long sequences, illustrating how important this basic mechanism has become for the sustenance of life.
\begin{acknowledgments}
We thank Eric Galburt, Justin Bois and Abigal Klopper for fruitful discussions and suggestions. JMRP acknowledges financial support from grants MOSAICO (Spanish Government) and MODELICO (Comunidad de Madrid). SWG acknowledges funding by the EMBO young investigator program and the Paul Ehrlich Foundation. MD acknowledges partial support from FOM, which is financially supported by the ``Nederlandse Organisatie voor Wetenschappelijk Onderzoek". This research was supported in part by the National Science Foundation under Grant No. NSF PHY05-51164.
\end{acknowledgments}

\end{document}